\documentclass[12pt]{article}
\usepackage{a4wide}
\usepackage{latexsym}
\usepackage{amsmath}
\usepackage{amsfonts}
\usepackage{amscd}
\usepackage{cite}
\usepackage{graphicx}
\usepackage{float}

\usepackage{pslatex}
\usepackage[latin1]{inputenc}
\usepackage[OT2,T1]{fontenc}

\newcommand{\bq}{\begin{eqnarray}}
\newcommand{\eq}{\end{eqnarray}}
\newcommand{\eps}{\varepsilon}

\DeclareSymbolFont{cyrletters}{OT2}{wncyr}{m}{n}
\DeclareMathSymbol{\Sha}{\mathalpha}{cyrletters}{"58}

\usepackage{color}

\begin{document}

\thispagestyle{empty}

\begin{flushright}
  MZ-TH/12-39
\end{flushright}

\vspace{1.5cm}

\begin{center}
  {\Large\bf Direct contour deformation with arbitrary masses in the loop \\
  }
  \vspace{1cm}
  {\large Sebastian Becker and Stefan Weinzierl\\
\vspace{2mm}
      {\small \em PRISMA Cluster of Excellence, Institut f{\"u}r Physik, }\\
      {\small \em Johannes Gutenberg-Universit{\"a}t Mainz,}\\
      {\small \em D - 55099 Mainz, Germany}\\
  } 
\end{center}

\vspace{2cm}

\begin{abstract}\noindent
  {
We present a method, which constructs a suitable deformation vector in loop momentum space, when the loop integration is done
numerically with the help of the subtraction method.
The method presented here extends previously discussed techniques from the massless case to the general case of arbitrary masses in the loop.
   }
\end{abstract}

\vspace*{\fill}

\newpage

\section{Introduction}

NLO predictions for multi-jet final states are important for the LHC experiments.
In recent years there has been tremendous progress in our ability to compute observables at NLO with a sizable number of
final state partons. NLO calculations with six or seven final state partons are now feasible \cite{Berger:2010zx,Ita:2011wn,Becker:2011vg}.
One of the techniques which allow to go to such high parton multiplicities is a purely numerical approach.
Within this approach also the virtual one-loop corrections are treated by Monte Carlo methods.
In the numerical approach the NLO contribution to an observable is given as the sum of three sub-contributions.
The first one is the real emission part minus 
suitable and well-known subtraction terms \cite{Catani:1997vz,Dittmaier:1999mb,Phaf:2001gc,Catani:2002hc,Dittmaier:2008md,Czakon:2009ss,Gotz:2012zz}.
The subtraction terms ensure that the phase space integral can be performed in four space-time dimensions with Monte Carlo techniques.
The computation of this part is similar for all groups working in the field of multi-parton NLO calculations.

The techniques of the various groups differ how they compute the virtual part. 
Within the numerical approach the ideas of the subtraction method and Monte Carlo integration are carried over to the virtual part.
One subtracts from the one-loop amplitude suitable approximation terms for the soft, collinear and 
ultraviolet singularities \cite{Becker:2012aq,Becker:2010ng,Assadsolimani:2010ka,Assadsolimani:2009cz,Nagy:2003qn}.
The difference is integrable and the integration over the loop momentum can be combined with the integration over the phase space of the final state
particles in one Monte Carlo integration.

The third contribution to an NLO observable is given by the integrated subtraction terms from the real and from the virtual part.
The sum of the real and virtual subtraction terms is finite and the phase space integration can again be performed by Monte Carlo methods.
From a computational point of view this contribution is only slightly more expensive than the corresponding Born contribution.

The essential part of the numerical method is the treatment of the virtual corrections.
Apart from the subtraction terms for the integrand of the one-loop amplitude the second important ingredient is a method for the contour deformation.
In the vicinity of a singularity we use, whenever possible, contour deformation to by-pass the singularity.
In the case where this is not possible, because the contour is pinched, and if the singularity is not integrable
then (and only then) there is a subtraction term for it.
The quality of the choice of the integration contour directly translates into the final Monte Carlo integration error.

Several algorithmic approaches for the contour deformation have been discussed in the
literature \cite{Becker:2010ng,Assadsolimani:2010ka,Assadsolimani:2009cz,Gong:2008ww,Borowka:2012yc,Anastasiou:2007qb,Nagy:2006xy,Nagy:2003qn,Soper:2001hu,Soper:1999xk,Soper:1998ye}.
They can be classified roughly according to three categories: In the first category one introduces Feynman parameters and integrates (after Feynman integration) over
the loop momentum analytically. Contour deformation (possibly in combination with sector decomposition) is applied to the integration over the Feynman parameters.
Also in the second category Feynman parameters are introduced, however the loop integration is not done analytically but numerically.
Contour deformation is applied to the integration over the loop momentum and the integration over the Feynman parameters.
Within the third category no Feynman parameters are introduced and contour deformation is applied to the integration over the loop momentum.
This is called ``direct contour deformation''.
An algorithm to construct the contour deformation within the third category has been given in ref.~\cite{Gong:2008ww} for the case where all internal particles are massless.

In our calculation for $e^+e^- \rightarrow 7 \;\mbox{jets}$ it turned out that the direct deformation method performs best.
In view of the LHC experiments, where massive particles like the Higgs, the top quark, the $W$- and $Z$-boson or hypothetical heavy particles beyond the Standard Model
play an important role, it is desirable to generalise the direct deformation method to arbitrary masses.
This is the purpose of the present paper.
We present an algorithm, which constructs the deformation vector in the general case. In particular we can have some internal masses non-zero and others equal to zero.
We assume that if an internal mass is non-zero then it is non-negligible against the centre-of-mass energy of the experiment.
This is the case for heavy particles like the Higgs, top, $W$- and $Z$-bosons at the LHC. 
It excludes the treatment of mass effects of extremely light particles like the electron in LHC experiments (which also for other reasons are non-trivial to treat numerically).

The generalisation to arbitrary masses is not a straightforward task. The method of ref.~\cite{Gong:2008ww} for the massless case is built on the fact that the singularities
are located on light-cones, which are nested inside each other.
Unfortunately the property that the singular surfaces are nested inside each other no longer holds if we replace the cones by more general hyperboloids.
We solve this problem with two techniques:
On the one hand we introduce additional reference points within the intersection of forward with backward mass hyperboloids.
On the other hand we probe pre-defined standard directions in order to check if a deformation in this direction is possible.

This paper is organised as follows:
In section~\ref{sect:numerical_integration} we define the problem and introduce our notation.
Section~\ref{sect:contour_deformation} is the main part of this paper and specifies the algorithm for the contour integration.
In section~\ref{sect:checks} we describe checks which we have performed to test our method.
Finally, section~\ref{sect:conclusions} contains our conclusions.

\section{Numerical integration of loop amplitudes}
\label{sect:numerical_integration}

In this paper we consider primitive one-loop amplitudes.
Primitive amplitudes are gauge-invariant building blocks of the full one-loop amplitude.
Each primitive amplitude has a fixed cyclic ordering of the external particles.
The cyclic ordering ensures that there are at maximum $n$ different loop propagators in the problem, where $n$ is the 
number of external legs.
We label the external momenta clockwise by $p_1$, $p_2$, ..., $p_n$ 
and define $q_i=p_1+p_2+...+p_i$.
The loop momenta are
\bq
 k_j & = & k - q_j, 
 \;\;\;
 q_j = \sum\limits_{l=1}^j p_l.
\eq
For convenience we set
\bq
 k_0 = k_n & \mbox{and} & q_0 = q_n.
\eq
Due to momentum conservation we actually have
\bq
 q_0 = q_n = 0.
\eq
Nevertheless we will use $q_0$ (or $q_n$), this makes the formulae more symmetric with respect to the indices.
\begin{figure}
\begin{center}
\includegraphics[bb= 180 510 440 720,width=0.45\textwidth]{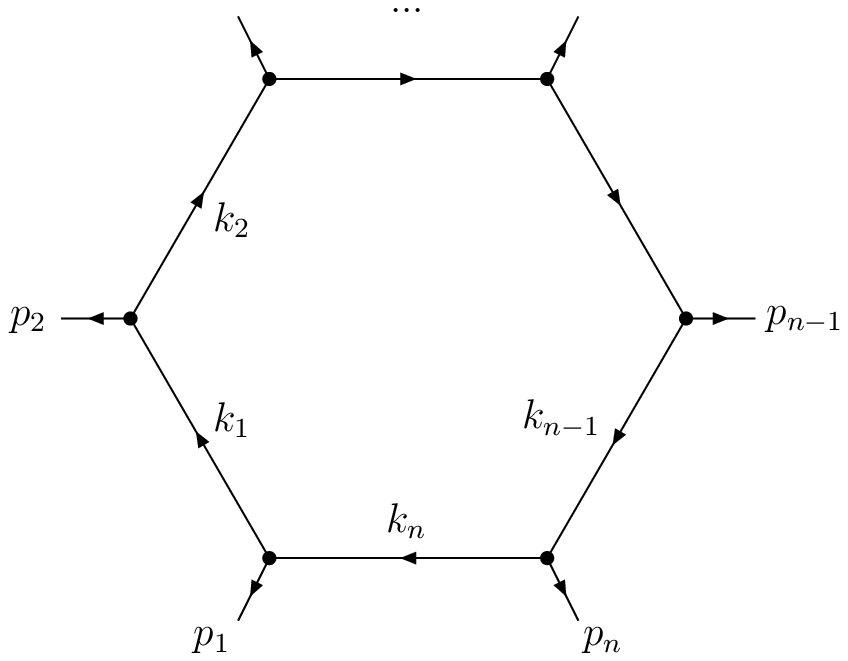}
\caption{\label{figure_momenta_one_loop}
The labelling of the momenta for a primitive one-loop amplitude. The arrows denote the momentum flow.}
\end{center}
\end{figure}
The labelling of the momenta is shown in fig.~(\ref{figure_momenta_one_loop}).

In addition we define an ultraviolet propagator with momentum
\bq
 \bar{k} & = & k - Q
\eq
and mass squared $\mu_{\mathrm{UV}}^2$.
The four-vector $Q$ and the parameter $\mu_{\mathrm{UV}}^2$ specify the ultraviolet propagator.
The mass squared $\mu_{\mathrm{UV}}^2$ can be chosen freely, and a useful choice is to take $\mu_{\mathrm{UV}}^2$ purely imaginary with
$\mbox{Im}\;\mu_{\mathrm{UV}}^2 <0$. Choosing the value along the negative imaginary axis large enough ensures that the ultraviolet
propagator never goes on-shell.

The object of investigation is an integral of the form
\bq
\label{complex_integral}
 I
 & = & 
 \int\frac{d^{4}k}{(2\pi)^{4}}
 \frac{R(k)}{\prod\limits_{j=1}^{n} \left( k_j^2 - m_j^2 \right)},
\eq
where $R(k)$ is a rational function of the loop momentum $k^{\mu}$, which has only poles at $\bar{k}^2-\mu_{\mathrm{UV}}^2=0$.
With a proper choice of $\mu_{\mathrm{UV}}^2$ one can ensure that the integration contour always stays away from the poles of the ultraviolet propagator.
Therefore the only poles which we have to care about are the ones which are shown
explicitly in eq.~(\ref{complex_integral}). 
It is assumed that the integral is finite, i.e. all contributions which would lead to poles in the dimensional parameter $\eps$ when the integral
is calculated in $D=4-2\eps$ dimensions are absent.
This can be achieved with the subtraction method.
Usually the integrand is the integrand of a primitive bare one-loop amplitude minus subtraction terms for the soft, collinear and ultraviolet
singularities.
Suitable subtraction terms can be found in the literature \cite{Assadsolimani:2009cz,Becker:2010ng,Becker:2012aq}.
Since the integral is finite, it can be computed in four dimensions.
However, this does not yet imply that we can safely integrate each of the four components 
of the loop momentum $k^\mu$
from minus infinity to plus infinity along the real axis.
There is still the possibility that some of the loop propagators go on-shell 
for real values of the loop momentum.
If the contour is not pinched this is harmless, as we may escape into the complex plane 
in a direction indicated by
Feynman's $+i\delta$-prescription.
However, it implies that the integration should be done over a region of real dimension $4$ 
in the complex space ${\mathbb C}^4$.
If the contour is pinched then the singularity is integrable when the integration is done over the loop momentum space and the phase space.
This is the case because either the singularity in the bare one-loop amplitude is integrable by itself, or -- if not -- there is a subtraction term
for it.

The purpose of this paper is to give a method for a suitable deformation into the complex space ${\mathbb C}^4$.
To this aim we set 
\bq
 k & = & 
 \tilde{k} + i \kappa(\tilde{k}),
\eq 
where $\tilde{k}^{\mu}$ is real. After this deformation our integral equals
\bq
\label{integral_after_deformation}
 I 
 & = &
 \int\frac{d^{4}\tilde{k}}{(2\pi)^{4}}
 \left|\frac{\partial k^{\mu}}{\partial \tilde{k}^{\nu}}\right|
 \frac{R(k(\tilde{k}))}{\prod\limits_{j=1}^{n} \left(\tilde{k}_{j}^{2}-m_{j}^{2}-\kappa^{2}+2i \tilde{k}_{j}\cdot\kappa \right)}.
\eq
The Jacobian 
\bq
 \left|\frac{\partial k^{\mu}}{\partial \tilde{k}^{\nu}}\right|
\eq
for the contour deformation can be computed numerically.
To match Feynman's $+i\delta$-prescription
we have to construct the deformation vector $\kappa$ such that
\bq
\tilde{k}_{j}^{2}-m_{j}^{2} & = & 0 \quad \rightarrow \quad \tilde{k}_{j}\cdot \kappa \geq 0,
\eq
and the equal sign applies only if the contour is pinched.
As long as $\kappa$ is infinitesimal we may neglect the term $\kappa^2$ in eq.~(\ref{integral_after_deformation}).
On the other hand in order to minimise Monte Carlo integration errors we would like to choose $\kappa$ as large as possible.
We will therefore first construct the direction of $\kappa$ and then scale $\kappa$ to the maximal length 
such that no poles are crossed by varying $\kappa$ from zero length to its maximal length.

The singularities of the integrand lie on mass hyperboloids defined by $(k-q_i)^2-m_i^2=0$
with origins given by $q_0$, $q_1$, ..., $q_{n-1}$.
\begin{figure}
\begin{center}
\includegraphics[bb=130 610 460 735]{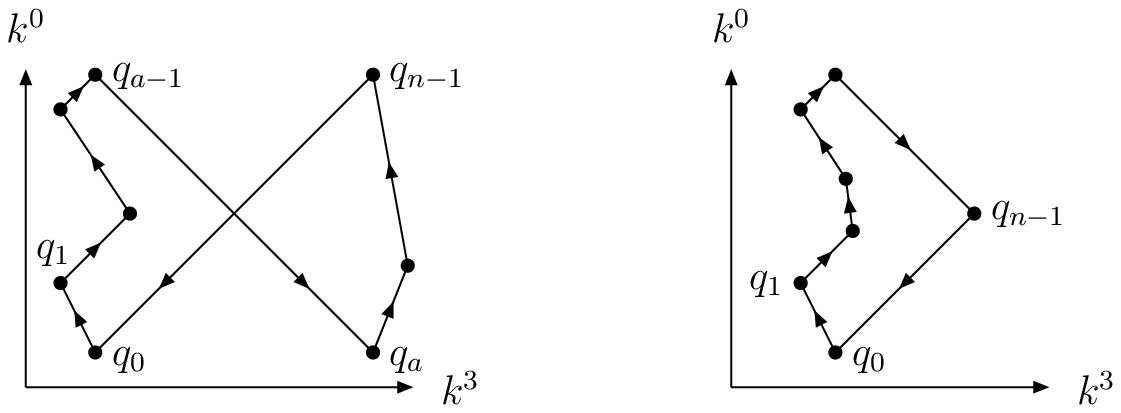}
\caption{\label{fig_zig_zag}
The origins of the mass hyperboloids in loop momentum space.
The diagram on the left shows the generic case for a 
primitive amplitude, the diagram on the right shows the degenerate case, where
the two incoming particles are adjacent.
}
\end{center}
\end{figure}
Since
\bq 
 p_i & = & q_i - q_{i-1},
\eq
the external momenta $p_i$ connect the origins of the mass hyperboloids and we arrive 
for a generic primitive one-loop amplitude at the
graphical representation shown in fig.~(\ref{fig_zig_zag}) on the left.
The lines connecting the points $q_i$ form a closed loop
due to momentum conservation
\bq
 \sum\limits_{i=1}^n p_i = 0.
\eq
Without loss of generality we may assume that one of the two momenta corresponding to the initial-state particles is $p_n$.
We then denote the momentum of the other initial-state particle by $p_a$.
In the case where the two initial-state particles are adjacent we can take $a=n-1$ and the diagram degenerates to the one shown
in fig.~(\ref{fig_zig_zag}) on the right.
In practice the two initial-state particles are always massless and the two initial-state momenta are light-like: $p_n^2=p_a^2=0$.

We define the forward mass hyperboloid as all points, which satisfy
\bq
 \left(k-q_i\right)^2 - m_i^2 = 0,
 & &
 k^0 \ge q_i^0.
\eq
Analogously, the backward mass hyperboloid is defined by
\bq
 \left(k-q_i\right)^2 - m_i^2 = 0,
 & &
 k^0 \le q_i^0.
\eq
The interior region of a mass hyperboloid is given by the set of all points, which satisfy
\bq
 \left(k-q_i\right)^2 - m_i^2 \ge 0.
\eq
We use the convention that the boundary is part of the interior region of a mass hyperboloid.

In loop momentum space the points $q_0$, $q_1$, ..., $q_{n-1}$ are contained within a finite region.
We call this region the ``interior region of the loop momentum space''. If it is clear from the context we will just use the term
``interior region'' for the interior region of the loop momentum space.
We may define the interior region of the loop momentum space by the set of all points, 
which are in the interior of some forward mass hyperboloid and in the interior of some backward mass
hyperboloid.
The exterior region of the loop momentum space is defined by the complement of the interior region of the loop momentum space. 
We note that only in the interior region intersections of forward hyperboloids with backward hyperboloids occur. 

\section{Contour deformation}
\label{sect:contour_deformation}

In this section we show how to construct the deformation vector $\kappa$.
We write
\bq
\kappa & = & \lambda\left(\kappa_{\mathrm{int}}+\kappa_{\mathrm{ext}}\right),
\eq
where $\kappa_{\mathrm{int}}$ is a deformation vector for the interior region of the loop momentum space, $\kappa_{\mathrm{ext}}$ is a deformation parameter for the exterior region  of the loop momentum space and
$\lambda$ is a scaling parameter.
We discuss the construction of the two contributions $\kappa_{\mathrm{ext}},\kappa_{\mathrm{int}}$ and the scaling parameter $\lambda$ separately.

Some functions in the construction below will depend on individual components of four-vectors and are therefore dependent on the Lorentz frame.
To make the definition unique we choose a specific Lorentz frame.
A possible choice is the centre-of-mass system of the event.
We make the convention that all frame-dependent statements refer to the centre-of-mass frame.

In order to make this section more readable we will write throughout this section for the real loop momentum $k$ instead of $\tilde{k}$.

\subsection{Helper functions}

We define some functions which are used in the construction of the deformation vector.
We start with two functions $h_{\delta+}$ and $h_{\delta-}$ defined by
\bq
\label{dd3}
h_{\delta\pm}(k,m^{2},M_1^2)&=&\frac{\left(\pm k^{0}-\sqrt{\vec{k}^{2}+m^{2}}\right)^{2}}{\left(\pm k^{0}-\sqrt{\vec{k}^{2}+m^{2}}\right)^{2}+M_{1}^{2}}.
\eq
The interpretation of the two functions $h_{\delta\pm}$ is as follows:
The function $h_{\delta +}$ vanishes whenever $k$ lies
on the forward mass hyperboloid defined by
\bq
k^{2}-m^{2}&=&0 \qquad \mbox{and}\qquad k^{0}\ \geq \ 0.
\eq
Analogously the function $h_{\delta -}$ vanishes whenever $k$ lies 
on the backward mass hyperboloid defined by
\bq
k^{2}-m^{2}&=&0 \qquad \mbox{and}\qquad k^{0}\ \leq \ 0.
\eq
If $k$ is far away from the hyperboloid the function converges to $h_{\delta\pm}=1$.
The parameter $M_1^2$ controls the size of the transition region, where the function increases from zero to one.
A typical value for the parameter $M_1^2$ is given by $M_1 = 0.035 \sqrt{\hat{s}}$, where $\sqrt{\hat{s}}$ is the centre-of-mass energy of the partonic event.

We further define a function $h_{\delta}$ by
\bq
h_{\delta}(k,m^{2},M_1^2)&=&  
\frac{\left(|k^{0}|-\sqrt{\vec{k}^{2}+m^{2}}\right)^{2}}{\left(|k^{0}|-\sqrt{\vec{k}^{2}+m^{2}}\right)^{2}+M_{1}^{2}}.
\eq
The function $h_{\delta}$ vanishes whenever $k$
lies on the (forward or backward) mass hyperboloid defined by
\bq
k^{2}-m^{2}&=&0,
\eq
and converges to one far away from the hyperboloid.
The functions $h_{\delta+}$ and $h_{\delta-}$ are similar to the functions $h_+$ and $h_-$ defined in \cite{Gong:2008ww} for the massless case, except that they do not contain
any Heaviside step function. As a consequence they only cut out a region near the forward or backward mass hyperboloid, respectively, but not the region inside
the mass hyperboloid.

We further define a function $h_{\theta}(t,M_1^2)$ by
\bq
 h_{\theta}\left(t,M_1^2\right)
 & = & 
 \frac{t}{t + M_1^2} \theta\left( t \right),
\eq
where $\theta(t)$ is the Heaviside step function. 
The function $h_{\theta}(t,M_1^2)$ vanishes whenever the variable $t$ is smaller zero
and converges to one for $t \gg 0$.

\subsection{The deformation in the exterior region}

We start with the definition of the deformation vector $\kappa_{\mathrm{ext}}$ for the exterior region.
In the exterior region we have per definition no intersections between forward and backward mass hyperboloids. 
Therefore we can separate the cases, where the loop momentum lies on a forward hyperboloid from the cases, where the loop momentum lies on a backward
hyperboloid.
Pinch singularities do not occur in the exterior region.
As a further requirement on the deformation vector in the exterior region we have to demand that the ultraviolet power counting is respected.
In other words we have to ensure that each propagator falls off like $k^{-2}$ for $k \rightarrow \infty$ independent of the direction in which we
approach infinity.

We write the deformation vector $\kappa_{\mathrm{ext}}$ as a sum of two terms.
The first term deforms correctly if the loop momentum lies on a forward hyperboloid, while the second term deforms correctly if the loop momentum 
lies on a backward hyperboloid.    
The deformation vector for the exterior region is given by
\bq
\kappa_{\mathrm{ext}}^{\mu}(k)&=&g_{\mu\nu}\left(c_{+}k^{\nu}_{+}+c_{-}k^{\nu}_{-}\right) 
\eq
with
\bq
\label{def_kplusminus}
k_{\pm}& = & k-P_{\pm}.
\eq
and $g_{\mu\nu}=\mbox{diag}(1,-1,-1,-1)$ the metric tensor.
The four-vectors $P_{\pm}$ are given by
\bq
P_{+}&=&Z_{+}(q_{a}+q_{0},q_{a}-q_{0}), \nonumber \\
P_{-}&=&Z_{-}(q_{a-1}+q_{n-1},q_{a-1}-q_{n-1}), 
\eq
with
\bq
Z_{\pm}^{\mu}(x,y)&=&\frac{1}{2}\left(x^{\mu}\pm\frac{y_{\nu}}{|\vec{y}|}\left(g^{0\mu}y^{\nu}-g^{0\nu}y^{\mu}\right)\right).
\eq
The four-vector $P_+$ lies on the intersection of the backward light-cone from $q_0$ with the backward light-cone from $q_a$ in the plane spanned by the four-vectors
$q_0$ and $q_a$.
\begin{figure}
\centering
\includegraphics[width=0.45\textwidth]{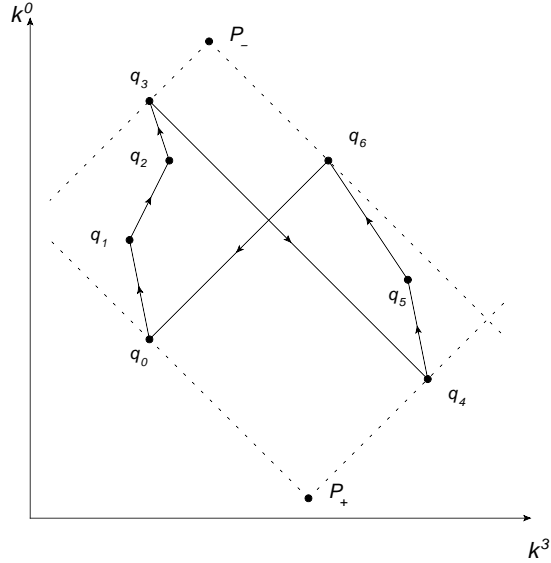}
\caption{\label{fig_Pplusminus} 
The definition of the four-vectors $P_{+}$ and $P_{-}$ in the loop momentum space for a primitive amplitude with $n=7$ and $a=4$.}
\end{figure}
The four-vector $P_-$ lies on the intersection of the forward light-cone from $q_{a-1}$ with the forward light-cone from $q_{n-1}$ in the plane spanned by the four-vectors
$q_{a-1}$ and $q_{n-1}$.
This is shown graphically in fig.~(\ref{fig_Pplusminus}).

The coefficients $c_\pm$ are defined as
\bq
c_{\pm} & = & \prod\limits_{i=1}^{n}h_{\delta\mp}\left(k_{i},m_{i}^{2},M_1^2\right).
\eq
If the loop momentum $k$ lies on a backward hyperboloid defined by
\bq
k_{i}^{2}-m_{i}^{2}&=&0\qquad \mbox{and}\qquad k_i^{0}\ \leq \ 0, \quad i\in\{1,\ldots,n\},
\eq
then one has $h_{\delta -}(k_i,m_i^2,M_1^2)=0$ and therefore $c_{+}=0$. 
Analogously, if the loop momentum $k$ lies on a forward hyperboloid given by
\bq
\label{case_exterior_forward_hyperboloid}
k_{i}^{2}-m_{i}^{2}&=&0 \qquad \mbox{and}\qquad k_i^{0}\ \geq \ 0,\quad i\in\{1,\ldots,n\}
\eq
then we have $h_{\delta +}(k_i,m_i^2,M_1^2)=0$ and therefore $c_{-}=0$. 
The coefficients $c_{\pm}$ ensure that all contributions vanish which would contribute with the wrong sign to the imaginary part.
It is easy to show that $\kappa_{\mathrm{ext}}$ always deforms correctly in the exterior region.
Let us assume again that the loop momentum $k$ lies on a forward hyperboloid as in eq.~(\ref{case_exterior_forward_hyperboloid}).
It follows that in this case
\bq
 k_i^0 & \ge & \left| \vec{k}_i \right|.
\eq
At the same time $k$ lies inside the forward cone with origin $P_+$ and we also have
\bq
 k_+^0 & \ge & \left| \vec{k}_+ \right|.
\eq
We further have in this case $c_-=0$ and $c_+ \approx 1$.
For the imaginary part we have to consider
\bq
 \kappa_{\mathrm{ext}} \cdot k_i & \approx & k_+ \circ k_i
 \ge \left| \vec{k}_+ \right| \left| \vec{k}_i \right| \left( 1 + \cos \theta_{+,i} \right)
 \ge 0.
\eq
Here we denoted by $\circ$ the Euclidean scalar product of two four-vectors and by $\theta_{+,i}$ the angle between the spatial three-vectors $\vec{k}_+$ and $\vec{k}_i$.  

We now discuss the ultraviolet behaviour of the deformation.
We note that $k_{i}^{2}$ can be small even if the loop momenta $k$ is in the ultraviolet region. 
We demand that the propagator falls off like $1/(k\circ k)$ in the ultraviolet region after contour deformation. 
In the exterior region at least one of the coefficients $c_{\pm}$ is unequal to zero. 
The scaling parameter $\lambda$ is of order $1$ in the ultraviolet region.
The imaginary part of a propagator in the ultraviolet region can be estimated by
\bq
k_{i}\cdot\kappa&\approx&k\cdot\kappa_{\mathrm{ext}}\ \approx \ k\circ k,
\eq
and therefore all propagators fall off like $1/(k\circ k)$ as desired.

\subsection{The deformation in the interior region}

We now consider the definition of the deformation vector $\kappa_{\mathrm{int}}$ for the interior region.
Here we face several situations:
In the interior region we have to take regions into account where a forward hyperboloid intersects with a backward hyperboloid. 
There can be regions which are bounded by more than two hyperboloids.
If massless particles are present we can also have cones which are tangential to each other. 
We split the deformation vector for the interior region into three parts. 
The first part is sufficient for the massless case, while the second part 
provides a deformation when two non-zero mass hyperboloids intersect.
Intersections of three or more mass hyperboloids are handled by the third part.
We write
\bq
\label{def_massive_interior}
\kappa_{\mathrm{int}}^{\mu}&=&-\sum\limits_{i=1}^{n}c_{i}k_{i}^{\mu}-\sum\limits_{\substack{i,j=1\\i<j}}^{n}c_{ij}k^{\mu}_{ij}
+ \kappa_{\mathrm{soft}}^{\mu}
\eq
with
\bq
k_{i}&=&k-q_{i} \qquad\mbox{and}\qquad k_{ij}\ = \ k-v_{ij}.
\eq
We note that if all particles are massless it is sufficient to define
\bq
\label{def_all_massless}
\kappa_{\mathrm{int}}^{\mu}&=&-\sum\limits_{i=1}^{n}c_{i}k_{i}^{\mu}.
\eq 
Up to minor modifications the definition in eq.~(\ref{def_all_massless}) corresponds to the one given in ref.~\cite{Gong:2008ww}.
However, if massive particles are present the extra terms in eq.~(\ref{def_massive_interior}) are required.
We start by giving the definitions of the coefficients 
$c_{i}$ and $c_{ij}$, then discuss the vectors $v_{ij}$ and specify $\kappa_{\mathrm{soft}}$ in the end.
The coefficients are defined such that they are zero if the corresponding vector deforms into the wrong direction. In formulae
\bq
k_{l}^{2}-m_{l}^{2} & = & 0 
 \quad\mbox{and}\quad
 \left\{\begin{array}{rclcrcl}
   -c_{i}k_{i}\cdot k_{l}&<&0&\Rightarrow&c_{i}&=&0, \\
   -c_{ij}k_{ij}\cdot k_{l}&<&0&\Rightarrow&c_{ij}&=&0, \\
 \end{array}\right.
 \quad
\forall l\in\{1,\ldots,n\}.
\eq
This ensures that the deformation vector $\kappa_{\mathrm{int}}$ never deforms in the wrong direction. 
We set
\bq
c_{i} & = & g\left(k_{\mathrm{centre}},\gamma_1,M_2^2\right) \prod\limits_{l=1}^{n}d_{i,l}
 \qquad\mbox{and}\qquad
c_{ij}\ =\ g\left(k_{\mathrm{centre}},\gamma_1,M_2^2\right) \prod\limits_{l=1}^{n}d_{ij,l},
\eq
with the function
\bq
g(k,\gamma_1,M_2^2)&=&\frac{\gamma_1 M_2^{2}}{k\circ k+M_2^{2}},
 \quad
 k_{\mathrm{centre}} = \frac{1}{2} \left( k_+ + k_- \right).
\eq
The momenta $k_+$ and $k_-$ have been defined in eq.~(\ref{def_kplusminus}).
The function $g$ ensures that the deformation vector $\kappa_{\mathrm{int}}$ falls off like $1/(k\circ k)$ in the ultraviolet region. 
Typical values for the parameters $\gamma_1$ and $M_2^2$ are $\gamma_1=0.7$ and $M_2 =0.7 \sqrt{\hat{s}}$.

The factors $d_{i,l}$ are defined by
\bq
\label{def_dil}
d_{i,l} & = &
 \left\{\begin{array}{rcl} 
  1 & : & l=i,\ m_{l}=0 \\
 [0.3cm]
 h_{\delta+}(k_{l},m_{l}^{2},M_1^2) & : & (q_{i}-q_{l})^{2}=0,\ q_{i}^{0} < q_{l}^{0},\ m_{l}=0 \\
 [0.3cm]
 h_{\delta-}(k_{l},m_{l}^{2},M_1^2) & : & (q_{i}-q_{l})^{2}=0,\ q_{i}^{0} > q_{l}^{0},\ m_{l}=0 \\
 [0.3cm]
 \max\left[h_{\delta}(k_{l},m_{l}^{2},M_1^2),h_{\theta}(-2k_{l} \cdot k_{i},M_1^2)\right] &:&\mbox{otherwise}   
 \end{array} \right.
 \nonumber \\
\eq
The factors $d_{ij,l}$ are defined by
\bq
\label{def_dijl}
d_{ij,l} & = & 
 h_{\theta}(z_{ij},M_1^2) \ \max\left[h_{\delta}(k_{l},m_{l}^{2},M_1^2),h_{\theta}(-2k_{l} \cdot k_{ij},M_1^2)\right],
 \nonumber \\
 z_{ij} & = & \left(q_i-q_j\right)^2 - \left(m_i+m_j\right)^2.
\eq   
Let us explain the idea behind these definitions. 
The first three lines in eq.~(\ref{def_dil}) handle exceptions due to massless particles in the loop.
In collinear configurations the contour is pinched in all directions perpendicular to the collinear axis, but not along the collinear axis.
The first three lines allow for the possibility to deform along the collinear axis.

In all other cases one considers the conditions 
$k_{l}^{2}=m_{l}^{2}$ and if $-k_{i}\cdot k_{l}<0$. If both conditions are fulfilled $d_{i,l}$ is zero.
This ensures that all terms contribute with the correct sign to the imaginary part.

The four-vectors $v_{ij}$ are defined by
\bq
 v_{ij} & = & \frac{1}{2} \left( q_i + q_j - \frac{(m_i-m_j)}{\sqrt{(q_i-q_j)^2}} (q_i-q_j) \right).
\eq
The geometric interpretation 
\begin{figure}
\centering
\includegraphics[width=0.45\textwidth]{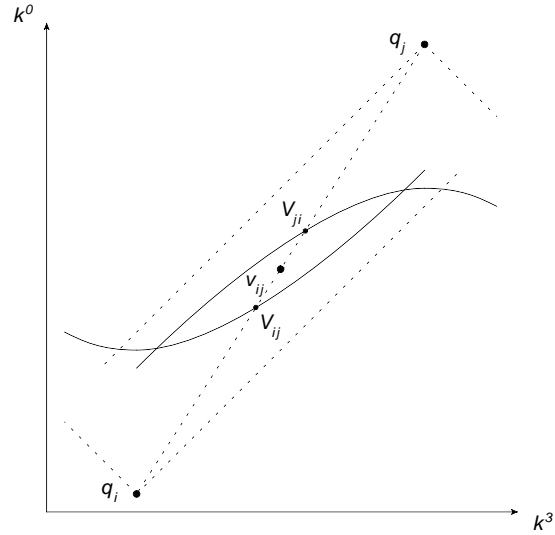}
\caption{\label{fig_vij} 
Definition of the vector $v_{ij}$. The line connecting  the points $q_{i}$ and $q_{j}$ cuts the mass hyperboloids at the two points $V_{ij}$ and $V_{ji}$. The vector $v_{ij}$ is at half of the distance between these two cutting points.}
\end{figure} 
of the vector $v_{ij}$ is shown in fig.~(\ref{fig_vij}).

If three or more mass hyperboloids intersect, it can happen that none of the vectors $v_{ij}$ lies inside the intersection region.
\begin{figure}
\centering
\includegraphics[width=0.45\textwidth]{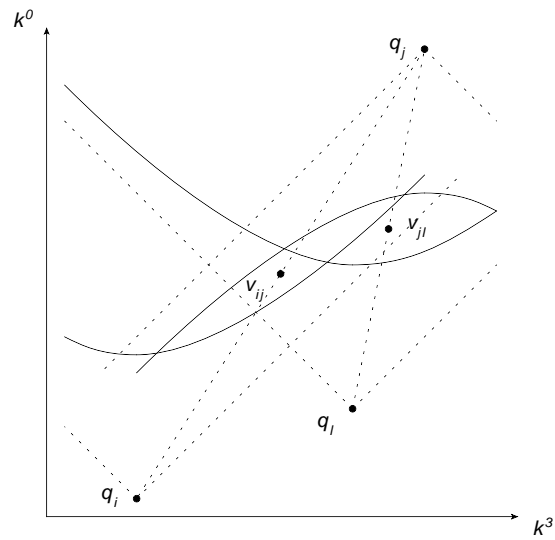}
\caption{\label{fig_empty_intersection} 
If three or more mass hyperboloids intersect, the vector $v_{ij}$ need not lie inside the intersection region.
}
\end{figure} 
An example of this situation is shown in fig.~(\ref{fig_empty_intersection}).
In this case the vectors defined up to now do not provide a deformation, since all coefficients vanish.
The full interior region has a typical length scale of $\sqrt{\hat{s}}$. Regions, where three or more mass hyperboloids intersect have typically
a much smaller length scale.
The vector $\kappa_{\mathrm{soft}}$ provides a deformation in these regions.
Since the algebraic formulae describing the intersection of three or more mass hyperboloids are already rather complicated to handle, we use
for the construction of $\kappa_{\mathrm{soft}}$ a different approach.
In order to find a suitable deformation vector we probe a pre-defined set of directions and accept a given direction, if the deformation goes in the right
direction. We write
\bq
 \kappa_{\mathrm{soft}}
 & = &
 \sum\limits_{a} c_a \kappa_a,
\eq
where the index $a$ sums over all pre-defined directions.
It turns out that the set of the four directions along the Cartesian coordinate axis does already a good job.
We set
\bq
 \kappa_0 = E_{\mathrm{soft}} \left(1,0,0,0\right), & & \kappa_1 = E_{\mathrm{soft}} \left( 0,1,0,0\right),
 \nonumber \\
 \kappa_2 = E_{\mathrm{soft}} \left(0,0,1,0\right), & & \kappa_3 = E_{\mathrm{soft}} \left( 0,0,0,1\right),
\eq
with $E_{\mathrm{soft}}$ being an energy scale much smaller then the centre-of-mass scale. A typical value is $E_{\mathrm{soft}}= 0.03 \sqrt{\hat{s}}$.
The coefficients $c^a$ are given by
\bq
 c_a & = & 
 g\left(k_{\mathrm{centre}},\gamma_1,M_2^2\right) \left( \prod\limits_{l=1}^{n} d^+_{a,l} - \prod\limits_{l=1}^{n} d^-_{a,l} \right)
\eq
with 
\bq
\label{def_d_plus_minus}
 d^+_{a,l} & = &
 \max\left[h_{\delta}(k_{l},m_{l}^{2},\gamma_2 M_1^2),h_{\theta}(2k_{l} \cdot \kappa^a,\gamma_2 M_1^2)\right],
 \nonumber \\
 d^-_{a,l} & = &
 \max\left[h_{\delta}(k_{l},m_{l}^{2},\gamma_2 M_1^2),h_{\theta}(-2k_{l} \cdot \kappa^a,\gamma_2 M_1^2)\right].
\eq
The product of the $d^+_{a,l}$'s probes, if a deformation in the direction of $\kappa_a$ is allowed, while the product
of the $d^-_{a,l}$'s probes if a deformation in the direction of $(-\kappa_a)$ is allowed.
Far away from all mass shells all functions $d^\pm_{a,l}$ are one and the contribution from $\kappa_a$ vanishes.
If we approach a mass shell, the values of the functions $d^\pm_{a,l}$ start to differ, thus yielding the right deformation vector.
Note that the factor $\gamma_2$ in eq.~(\ref{def_d_plus_minus}) ensures that also the transition region for the functions $h_\delta$ and $h_\theta$
has a smaller scale.  
A typical value for $\gamma_2$ is $\gamma_2=0.008$.

\subsection{The scaling parameter $\lambda$}

Having defined the deformation vectors $\kappa_{\mathrm{ext}}$ and $\kappa_{\mathrm{int}}$ for the exterior and interior region, respectively we now
discuss the definition of the scaling parameter $\lambda$ entering the definition of the full deformation vector $\kappa$:
\bq
\kappa & = & \lambda\left(\kappa_{\mathrm{int}}+\kappa_{\mathrm{ext}}\right).
\eq
For the sum of $\kappa_{\mathrm{int}}$ and $\kappa_{\mathrm{ext}}$ we write
\bq
 \kappa_0 & = & \kappa_{\mathrm{int}} + \kappa_{\mathrm{ext}}.
\eq
The definition of the scaling parameter $\lambda$ is along the lines of \cite{Gong:2008ww} with some minor changes due to the masses 
which can appear in the loop propagators.

In the previous sections we discussed the construction of $\kappa_{0}=\kappa_{\mathrm{int}}+\kappa_{\mathrm{ext}}$ in detail. 
By construction $\kappa_0$ deforms in the right direction.
It follows immediately that if $\lambda$ is infinitesimal we have a correct deformation. 
But for small Monte Carlo errors we would like to deform as far as possible away from the singularities 
and therefore we need to make $\lambda$ as large as possible. 
In doing so we have to ensure that we do not cross any poles by varying the size of $\lambda$ from zero to its final value. 

We define for each propagator in the loop a scaling parameter $\lambda_{j}$ such that we do not cross any pole of the given propagator by varying $\lambda_{j}$ 
from zero to its final value. 
By taking $\lambda$ as the minimum of these $\lambda_{j}$ we ensure that we do not cross any poles. We write for the $j$-th propagator
\bq
D_{j}&=&\left(k_{j}+ i \lambda\kappa_{0}\right)^{2}-m_{j}^{2}
\ =\ k_{j}^{2}+2 i \lambda\kappa_{0}\cdot k_{j}-\lambda^{2}\kappa_{0}^{2}-m_{j}^{2}
\eq
The function $D_{j}$ vanishes for values of $\lambda$ given by
\bq
\lambda & = & i \frac{\kappa_0 \cdot k_j}{\kappa_0^2} \pm\sqrt{Y_{j}-X_{j}},
\eq
where we introduced the functions
\bq
\label{def_Xj_Yj}
 X_{j} 
 & = &
 \left(\frac{\kappa_{0}\cdot k_{j}}{\kappa_{0}^{2}}\right)^{2}
 \quad\mbox{and}\quad 
 Y_{j}\ = \ \frac{k_{j}^{2}-m_{j}^{2}}{\kappa_{0}^{2}}.
\eq
If $Y_{j}>X_{j}$ and $X_{j}\to 0$ we have a pole at
\bq
\lambda &=& \sqrt{Y_{j}}.
\eq
To avoid this pole we limit the value of $\lambda$ to one half of this value. 
If instead $Y_{j}<X_{j}$ the poles of the propagator $D_{j}$ are on the imaginary $\lambda$-axis 
and we can choose the real value of $\lambda$ as large as we like. 
We define 
\bq
\label{def_lambda_j}
\lambda_{j}^{2}&=&\left\{\begin{array}{rcl}
Y_{j}/4&:& 2X_{j}\ < \ Y_{j} \\
X_{j}-Y_{j}/4&:&0\ <\ Y_{j}\ <\ 2X_{j} \\
X_{j}-Y_{j}/2&:& Y_{j}\ <\ 0\end{array}\right.
\eq
Next we define a value $\lambda_{\mathrm{coll}}$ by
\bq
\lambda_{\mathrm{coll}}&=&\frac{1}{4C},
\eq
with
\bq
C&=&\sum\limits_{i=1}^{n}c_{i}+\sum\limits_{\substack{i,j=1\\i<j}}^{n}c_{ij}
 + \sum\limits_a \left| c_a \right|.
\eq
$\lambda_{\mathrm{coll}}$ has two functions: First of all, in the case where several coefficients from the set of the $c_i$ and $c_{ij}$
are non-zero, the scaling parameter $\lambda_{\mathrm{coll}}$ acts as an averaging procedure.
Secondly, in the collinear case the numerator and the denominator in the definition of $\lambda_j$ in eq.~(\ref{def_lambda_j}) goes to zero
and $\lambda_{\mathrm{coll}}$ ensures numerical stability in this limit.

In addition we define a parameter $\lambda_{\mathrm{UV}}$ by
\bq
\lambda_{\mathrm{UV}}
 & = &
\left\{\begin{array}{rcl}
1&:& 4\kappa_{0}\cdot\bar{k}\  > \ \operatorname{Im}(\mu_{\mathrm{UV}}^{2}) \\
\frac{\operatorname{Im}(\mu_{\mathrm{UV}}^{2})}{4\kappa_{0}\cdot\bar{k}}&:&  4\kappa_{0}\cdot\bar{k}\ \leq \ \operatorname{Im}(\mu_{\mathrm{UV}}^{2})
\end{array}\right.
\eq
$\lambda_{\mathrm{UV}}$ ensures that the imaginary part of the ultraviolet propagator is always positive.

The scaling parameter $\lambda$ is then given by
\bq
\lambda&=& \min\left[1,\lambda_{1},...,\lambda_{n},\lambda_{\mathrm{coll}},\lambda_{\mathrm{UV}}\right].
\eq

\section{Checks}
\label{sect:checks}

In order to test our method we verify that well-known one-loop integrals are computed correctly.
We fix $n$ external momenta $p_1$, $p_2$, ..., $p_n$ and $n$ internal masses $m_1$, $m_2$, ..., $m_n$.
We let $S_r$ be a subset of $\{1,...,n\}$ containing $r$ indices and we consider the integral
\bq
\label{check_integral_1}
 I
 & = & 
 \int\frac{d^{4}k}{(2\pi)^{4}}
 \frac{\prod\limits_{l\in S_r} \left( k_l^2 - m_l^2 \right)}{\prod\limits_{j=1}^{n} \left( k_j^2 - m_j^2 \right)}.
\eq
The integral is a scalar one-loop $(n-r)$-point function
\bq
 I
 & = & 
 \int\frac{d^{4}k}{(2\pi)^{4}}
 \frac{1}{\prod\limits_{\substack{j=1\\j\notin S_r}}^{n} \left( k_j^2 - m_j^2 \right)},
\eq
and we consider scalar one-loop $(n-r)$-point functions which are infrared and ultraviolet finite.
We compare our results with the results obtained from the program ``LoopTools'' \cite{Hahn:1998yk}.
The contour deformation is obtained from the $n$ propagators in the denominator of eq.~(\ref{check_integral_1})
and independent of the numerator in eq.~(\ref{check_integral_1}).
\begin{figure}
\centering
\includegraphics{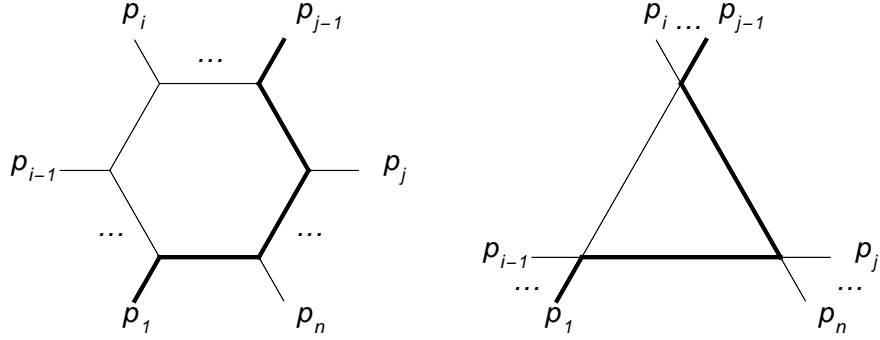}
\caption{\label{fig_pinching}
The diagram on the left defines the contour deformation for a process with $n-2$ massless and $2$ massive external legs. 
Pinching $(n-3)$ loop propagators results in a three point function shown on the right.}
\end{figure}
Although the analytical result is given as a $(n-r)$-point function, the contour deformation corresponds to a $n$-point function.
This is illustrated in fig.~(\ref{fig_pinching}).
By varying the set $S_r$ in the numerator we can probe different regions in loop momentum space.

We have verified the correctness of our method for randomly chosen external momenta and various choices of the internal masses.
We have found good agreement with the known analytical results at the per cent to the per mille level for a typical Monte Carlo integration
with $20 \cdot 10^6$ evaluations.
\begin{figure}
\begin{center}
\includegraphics[bb= 125 460 490 710,width=0.45\textwidth]{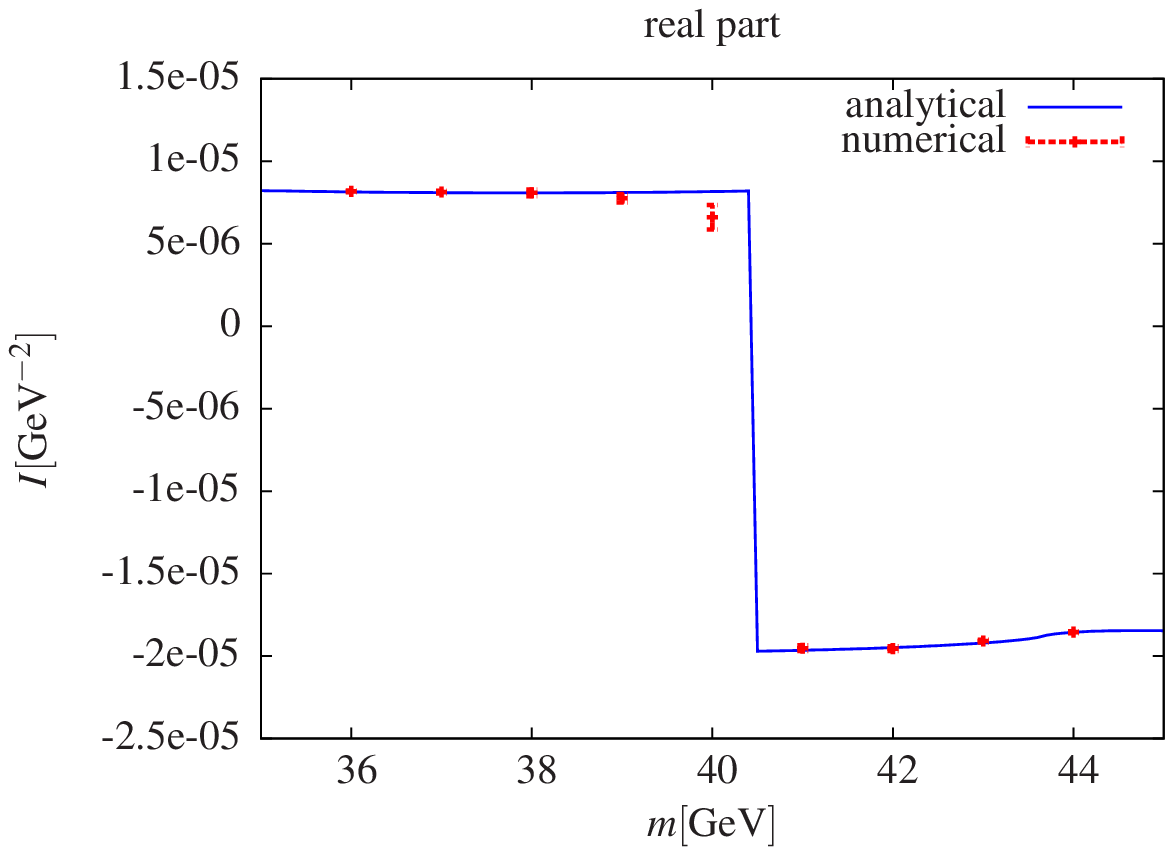}
\includegraphics[bb= 125 460 490 710,width=0.45\textwidth]{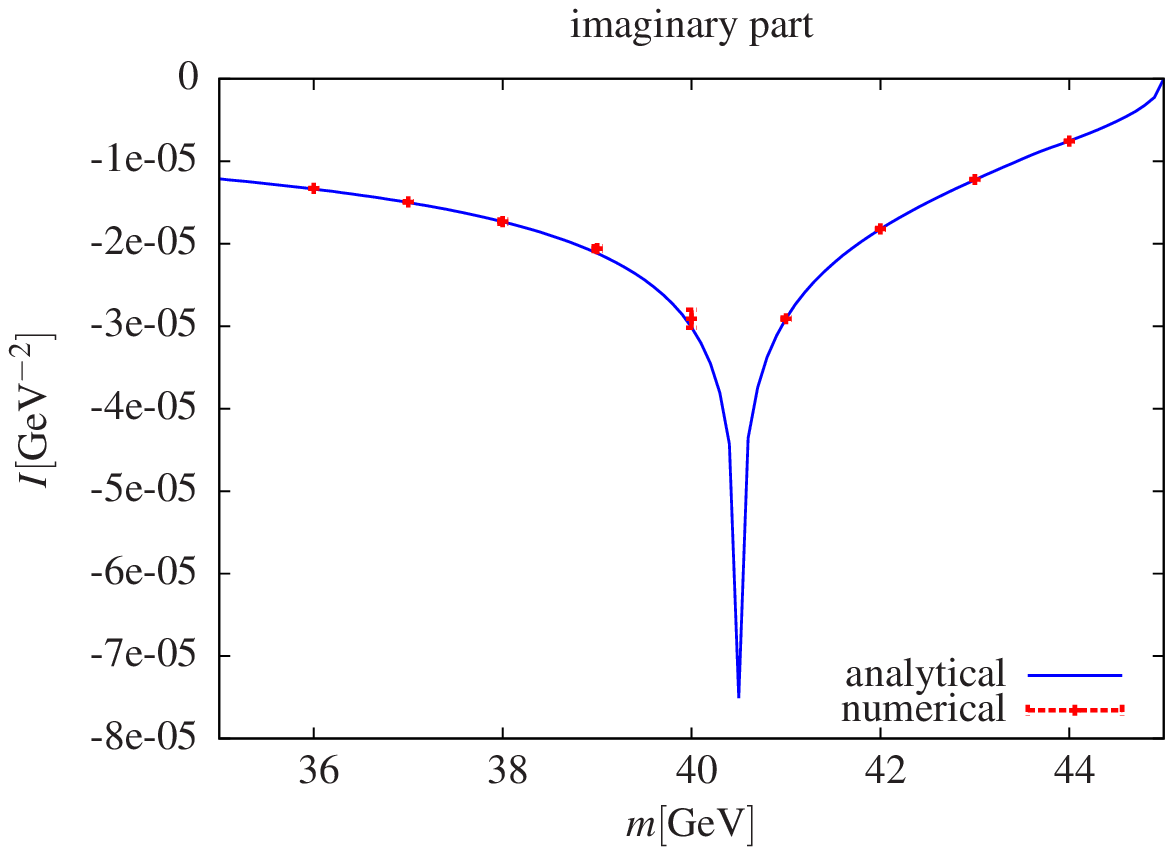}
\end{center}
\caption{\label{fig_threshold}
Comparison of the results obtained by Monte Carlo integration with the analytical results in the vicinity of a threshold.
}
\end{figure}
The most challenging parts from a numerical point of view are thresholds.
For example, one-loop functions may contain terms of the form
\bq
 \ln^2\left(-s+m^2\right).
\eq
Terms like these will give at the threshold $s=m^2$ discontinuities and logarithmic divergences.
In fig.~(\ref{fig_threshold}) we show a typical situation, where the real part exhibits a discontinuity while the imaginary part shows a divergence.
The analytical results have been obtained with ``LoopTools'' \cite{Hahn:1998yk}, while the numerical results have been obtained by Monte Carlo integration
with the method described in this paper.
The figure clearly shows, that also in the vicinity of thresholds the numerical method gives reasonable results.
The deformation defined by $\kappa_{\mathrm{soft}}$ plays an essential part in the vicinity of a threshold.


\section{Conclusions}
\label{sect:conclusions}

The numerical calculation of the virtual corrections rely on a method for the contour deformation.
In this paper we have given an algorithm to construct the deformation vector in loop momentum space for arbitrary masses in the loop.
The algorithm presented in this paper opens the possibility to treat massive particles within the direct deformation method.

\bibliography{/home/stefanw/notes/biblio}
\bibliographystyle{/home/stefanw/latex-style/h-physrev5}

\end{document}